# The Classification of Short and Long-term Driving Behavior for an Advanced Driver Assistance System by Analyzing Bidirectional Driving Features

Dr. Mudasser Seraj, Department of Civil and Environmental Engineering, University of Alberta

*Abstract*— **Insight into individual driving behavior and habits is essential in traffic operation, safety, and energy management. With Connected Vehicle (CV) technology aiming to address all three of these, the identification of driving patterns is a necessary component in the design of personalized Advanced Driver Assistance Systems (ADAS) for CVs. Our study aims to address this need by taking a unique approach to analyzing bidirectional (i.e. longitudinal and lateral) control features of drivers, using a simple rule-based classification process to group their driving behaviors and habits. We have analyzed high resolution driving data from the real-world CV-testbed, Safety Pilot Model Deployment, in Ann Arbor, Michigan, to identify diverse driving behavior on freeway, arterial, and ramp road types. Using three vehicular features known as jerk, leading headway, and yaw rate, driving characteristics are classified into two groups (Safe Driving and Hostile Driving) on short-term classification, and drivers' habits are categorized into three classes (Calm Driver, Rational Driver, and Aggressive Driver). Proposed classification models are tested on unclassified datasets to validate the model conviction regarding speeding and steep acceleration. Through the proposed method, behavior classification has been successfully identified in 86.31 ± 9.84% of speeding and 87.92 ± 10.04% of acute acceleration instances. In addition, our study advances an ADAS interface that interacts with drivers' in real-time in order to transform information about driving behaviors and habits into feedback to individual drivers. We propose an adaptive and flexible classification approach to identify both short-term and long-term driving behavior from naturalistic driving data to identify and, eventually, communicate adverse driving behavioral patterns.**

## I. INTRODUCTION

THE classification of individual driving behavior has played a vital role in identifying hazardous driving patterns, vehicle fuel consumption optimization, individualized vehicle control system design, and power management system design. Gradual expansion and integration of connected and autonomous vehicle (CAV) -based transportation systems have amplified the need to understand drivers' individual behaviors. Recognition and classification of driving behavior is now seen as intrinsic to the proper design and assessment of an Advanced Driver Assistance System (ADAS) as well as the enhancement of traffic safety through CAVs [1]–[6]. However, observations of real-world driving indicate that driving behavior is the result of instantaneous decisions made in response to the exogenous environment, including elements such as road type, surrounding traffic, and the physical and mental state of the driver. Assuming that these instantaneous driving decisions result from a complex fusion of different factors, this study aims to dynamically identify distinct types of driving behavior by analyzing bidirectional control, driver decisions. Developing a flexible yet accurate classifying process to identify both driving behavior and habits of individual drivers is our primary objective here.

Driving behavior is a complex concept, and the common association of 'Driving behavior' with 'Driving Habit/Style' complicates its definition and identification further. The correlation between the terms, as understood from the related literature, offers clarification on the distinct levels of classification. Driving behavior focuses exclusively on drivers' instantaneous decisions and correlates with the driving conditions experienced by drivers. Therefore, a precise understanding of the environment can provide better insight into driving behavior [7]–[9]. Furthermore, we can expect variations in decisions by same driver at different times for the same driving conditions because of transformed habitual influence. On the other hand, individual drivers' preferential driving behavior accumulates over time and develops into driving habit or driving style [1], [10]–[13]. While driving behavior varies with external factors, often erratically, driving habits change steadily in the longer term. The concepts of driving behavior and driving habit are necessary to distinguish between observed driving behavior on any given trip and developed driving habit from an accumulated driving history.

In this paper, we present a simplified approach to dynamically identifying driving behavior by analyzing drivers' jerk, yaw rate, and leading headway profiles on different roadways. Jerk, yaw rate, and leading headway profiles are regarded as indicators of individual drivers' longitudinal and lateral control decisions. Using these indicators, our research aims to decisively classify the behavior of any given driver. In so doing, we aim to contribute to driving behavior research in two ways: 1), we can generate more accurate representations that better identify hazardous driving behavior by analyzing bidirectional driving features for classification, and 2) we can establish and distinguish between the two different behavioral classes for individual trip behavior and accumulated driving history. Additionally, this paper presents our model for a convenient and cohesive ADAS interface that warns drivers in real time of unsafe driving behavior. This interface will also allow both drivers and regulatory organizations to review driving habits based on an accumulation of previous driving behavior.

The greater demand for understanding CAV justifies the need for comprehensive research on driving behavior. Traffic management authorities that apply the recommended approach derived from the results of this study could provide an efficient ADAS application of this promising technology to improve traffic mobility and safety. As such, the key contributions of this research are listed below:

i. The first contribution of this research is our account of both longitudinal and lateral driving features to detect adverse driving patterns from real-world driving data

ii. Another contribution of this research is the capture of the behavioral evolution of a driver's instantaneous responses (i.e. short-term behavior) to driving habit (i.e. long-term behavior)

iii. Finally, the resulting classification models are intended to propose a simple yet informative ADAS system to communicate detected behavioral information to drivers.

In order to best present our findings, the paper is organized as follows: Section II summarizes the leading literature on driver behavior classification, identifies the gaps on current knowledge, and outlines the contributions this research will make in order to address those gaps. Following on from this, Section III describes the proposed classification method in detail. Then, Section IV evaluates the proposed method's performance when identifying behavioral pattern, followed by a description of the plans to extend the current research. Finally, a synopsis of the research findings concludes this paper.

## II. LITERATURE REVIEW

Identification of driving behavior and habit has long been of interest among the researchers, especially with respect to enhancing road safety. Gibson and Crooks [14] conducted an early study on driving psychology and concluded that driving predominantly depends on drivers' perceptions of their surrounding environment. In particular, safe driving depends on a driver's psychological safe spatial zone. In other studies, different physical measures have been identified to capture drivers' perceptual decisions during driving [15]–[24]. Some specific driving measures included speeding and/or hard braking [25]–[31], jerky driving [1], [32]–[34], tailgating [35], [36], lane choice [37], [38], steering angle [29], [35], [38], [39], lateral acceleration [26], and the passing gap during overtaking [40]. These identified measures included both longitudinal and lateral features of driving to help categorize driving behavior. While speed and acceleration are frequently used measures of driving behavior, jerk profile is found to be more sensitive to safety-critical driving behavior [1]. With regards to longitudinal control, decision time and/or space headway are found to be more specific than speed, acceleration, or jerk profiles in reflecting hostile driving [35]. Since consistent headway during driving is the socially accepted norm of a safe driver, the choice of short and erratic headways could be explained in part by aggressive intentions. On the other hand, lateral control behavior is often associated with steering angle, lateral acceleration, or lane choice. Increased variations in these features can differentiate between safe and unsafe driving. While both longitudinal and lateral driving features play a role in defining driving behavior, the collective impact of both aspects remains uncharted.

A major motivation for the study of driving behavior identification is the development of techniques to modify driving behavior [41]–[47]. Recently, personalized communication through a connected vehicle (CV)-based ADAS system has been applied to driving behavior modification [48]–[51]. Driver identification that incorporates both driving behavior and driving habit is necessary to design an ADAS system that accounts for drivers' requirements, acceptability, and preferences [52]. However, labelling driving behavior from collected driving features varies widely in the literature. Major labelling techniques include rule-based [1], [9], [19], [53], [54], fuzzy logic-based [3], [55]–[57], and machine learning methods-based [2], [8], [12], [19], [28], [58], [59]. Due to their computational simplicity, robustness, and clear explanation, rule-based techniques of driving behavior identification are adopted by numerous studies. Larger set variables can create complex classification processes on rule-based methods that can then be solved by fuzzy logic-based methods. Due to the availability of large, multivariate datasets, machine learning methods have recently become prevalent among practitioners. While machine learning methods can identify driving patterns from big data with larger sets of variables, these labelling techniques often contain complex and delicate structures as well as inexplicable solutions. To avoid these pitfalls, we have used a small number of variables with relatively large datasets, and we have explored both rule-based and machine learning-based labelling techniques in order to choose an ideal technique for labelling unlabeled training data. Although the dataset chosen for this study may initially appear small, we feel that they are large enough to represent the behavioral variations of drivers as well as to demonstrate the proposed method of classification. Furthermore, the chosen dataset included the trips with a higher number of records than of the remaining trips in the dataset, which potentially accounts for most possible variations.

Reviewing the literature related to driving behavior identification, recognition, and classification, we noted that integration of bidirectional control decisions in classification could improve the odds of precise categorization, since the combination of both features can capture greater diversity potentially overlooked by one-dimensional feature-based classifications. Another key contribution of this study is to demonstrate the gradual development and changes in driving habits from driving behavior in both short-term and long-term driving behavior classifications. Finally, we have designed a user-friendly, real-time warning system for a driving behavior interface that includes the capability to provide long-term driving habit information and is a future extension of the current study.

## III. Methodology

In this study, we attempt to categorize both short term and long-term driving behavior. While the short-term classification represents a driver's individual trip behavior, the long-term classification stands for an individual driver's driving habit, formulated from previous driving experiences. Both classifications of driving behavior are based on a fixed duration (5 sec) moving window along the classification period. Short-term driving behavior is classified into two distinct classes, Safe Driving and Hostile Driving, as defined below:

- Safe Driving: driving instances within a trip when the driver anticipates the surrounding roadway environment and executes composed control decisions.
- Hostile Driving: driving instances within a trip when the driver fails to assess the surrounding roadway environment and compensates by performing impulsive and hazardous control decisions.

The continuous accumulation of short-term classifications, gathered from previous trips in the driving history, facilitates long-term driver behavior classification. In this classification process, individual drivers are grouped into three categories: Calm Driver, Rational Driver, and Aggressive Driver.

- Calm Driver: their share of cumulative hostile driving instances over the analysis period is below the specified lower threshold value
- Rational Driver: their share of cumulative hostile driving instances over the analysis period is within the lower and upper threshold value
- Aggressive Driver: their share of cumulative hostile driving instances over the analysis period is above the upper threshold value

### A. Data Preparation

The data used in this study was adopted from the Safety Pilot Model Deployment (SPMD) Project [60] database [61]. In this project real-world driving data were collected from roadways of Ann Arbor, Michigan through integrated safety devices and the radar-based data acquisition system that were developed by the Virginia Tech Transportation Institute [62]. These data were obtained via the Research Data Exchange website [63].. Sixty-three sensor-equipped vehicles were used to collect information from 13,792 trips containing 78.43 million data points at the frequency of 10Hz. Sensors attached to these CVs continuously collected information, including vehicle ID, trip ID, GPS longitude, GPS latitude, GPS UTC time, in-vehicle brake status, in-vehicle headlight status, in-vehicle speed, in-vehicle acceleration, in-vehicle steering position, in-vehicle throttle position, and in-vehicle yaw rate, amongst other types of information. From this large dataset, the top 550 trips (~4%), containing 7.94 million records in total (10.12%), were selected for our research by sorting the trips in descending order of available data records of each trip. Amongst the different operational data collected though the equipped vehicles' data acquisition systems, we chose three features for driving behavior classification: jerk, yaw rate, leading headway. In our analysis, these three features represented longitudinal and lateral control decisions undertaken by individual drivers.

As the first derivative of acceleration/deceleration and second derivative of velocity, jerk is a more effective feature than velocity or acceleration in driving behavior classification [1]. Also, longitudinal and lateral decisions of individual drivers are incorporated within this single feature. The jerk data for this study was measured from the vehicles' data acquisition system (i.e. Integrated Safety Device (ISD)) that recorded in-vehicle acceleration data at 10hz frequency. This main data file contained several fields detailing elements such as vehicle position and speed, fidelity measures of GPS-based data, and vehicle operation data (steering, throttles position etc.). The authors extracted the acceleration data from this large dataset to calculate jerk data.

Yaw rate measures the vehicle's lateral movement rate and characterizes a driver's lateral behavior. In contrast to earlier feature calculation, we obtained measurements of this feature directly from the vehicles' ISD system with the same frequency. Measurements of the leading headway stand for driver's longitudinal control decisions since the gap between vehicles often dictate car-following behavior. We collected leading headway data from the radar units that were installed as a part of vehicles' ISD units. These radars recorded the distance between the radar and the forward vehicle, in the cases where there was another vehicle within a 200 m distance in the same lane. The combination of these three mutually inclusive features - jerk, yaw rate, and leading headway - is capable of capturing instantaneous variations of drivers' bidirectional control

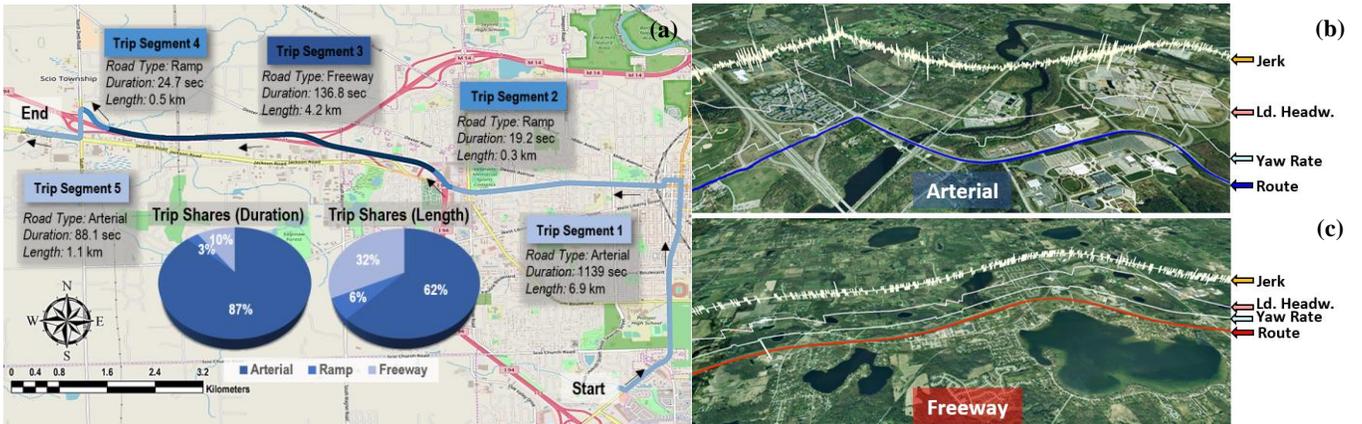

Fig. 1. (a) Road type-based segmentation of a sample trip, contrast of studied driving features on (b) arterial and (c) freeway.

decisions and, hence, assist in classifying drivers' behavior in real-time.

### B. Driver Behavior Classification Algorithm

The selected three features outlined above for classification were extracted from the chosen 550 trips. In addition to those three features, vehicle ID, trip ID, latitude, longitude, and time stamps were also included in the dataset, which we used to geographically locate the trip route and split the route based on road type [Figure 1(a)]. Figure 1(a) presented the segmentation of a sample trip from the dataset. Data points were placed on map from longitude and latitude information of the trip. The same information was used to classify the trip in different segments based on road type (i.e. arterial, ramp, freeway) traversed during the trip. Figure 1(a) shows three shades of blue that represent the three different classes of roads considered in this study, as well as details of the different segments. Two pie charts within the figure illustrate the proportion of trip duration and trip length for each class of road, from the whole trip. The assumption that one can observe substantial diversity in the driving environment between freeway and arterial roads motivated our road type-based splitting of trips. Since driving behavior is directly influenced by the surrounding environment, classifying driving behavior in relation to different road types using the same standards would lead to erroneous categorization. Additionally, visual observations of classifying features showed significant disparity between the road classes [Figure 1(b, c)]. Figure 1(b, c) highlights the distinctions between the different road type features for the trip that was plotted on Figure 1(a). Plotted feature profiles on arterial roads [Figure 1(b)] showed greater fluctuations of feature values in comparison to feature profiles on freeway [Figure 1(c)]. All three features showed relatively higher ranges of variability for arterials than freeways.

To emphasize driving behavior contrasts, each trip was divided into three road types, based on GPS location (i.e. longitude, latitude): Freeway, Arterial, and Ramp. Features of the same road types were grouped together to classify short-term and long-term driving behavior. From the training dataset (550 trips), 66.20% (5.26 million data points), 31.76% (2.52 million data points), and 2.04% (0.16 million data points) are labeled as freeway, arterial, and ramp respectively.

Once the features were sorted based on road types using the geolocation of each time stamp, their distribution was plotted

(**Figure 2**). We compared the datasets of each road type by an unpaired, two-sample t-test to justify our assumption of substantial feature disparity between road types. Comparison results of each pair (i.e. freeway vs arterial, arterial vs ramp, freeway vs ramp) presented significant differences (i.e. p-value < 0.0001) in the mean of each feature, with a 99% confidence level, while also assuming unequal variance of the tested samples.

Upon confirmation of attributional difference among road types, we calculated the absolute mean of each feature for the three road types, which was stored in a database. Next, we calculated the standard deviations of each feature for all trips with a moving window of $t_c = 5$ sec (50 data points). The coefficient of variation (CoV) was then calculated by dividing the measured standard deviations with the absolute mean of current road type within the time window (Equation 1). Since the CoV is the measure of relative variability, this statistical attribute of each driving feature was exerted when identifying hostile driving behavior for classification. Finally, we scaled the CoV datasets of each feature within [0 1] range for each of the road types (Equation 2). Since the absolute values of studied features were significantly different, the authors refrained from using the absolute values of these features and rather used scaled (i.e. standardized) coefficient of variations to perform classification.

$$CoV_f(t) = \frac{SD_f(t - t_c, \ t)}{\bar{f}_R} \tag{1}$$

$$CoV_f'(t) = \frac{CoV_f(t) - CoV_{f,R}^{min}}{CoV_{f,R}^{max} - CoV_{f,R}^{min}} \tag{2}$$

Here, $CoV_f(t)$ = coefficent of variation of feature $f$ (i.e. jerk, leading headway, yaw rate) at time $t$; $SD_f(t - t_c, \ t)$ = standard deviation of feature f within time $t - t_c$ and $t$ ($t_c = 5sec$); $\bar{f}_R$ = mean of absolute values of feature $f$ at roadtype $R$; $CoV_f'(t)$ = scaled coefficient of variation of feature $f$ at time $t$; $CoV_{f,R}^{min}$ = minimum coefficient of variation for feature $f$ on current roadtype $R$; $CoV_{f,R}^{max}$ = maximum coefficient of variation for feature $f$ at current roadtype $R$.

Once scaled, and the unlabeled CoVs of features were available, we were able to explore the labeling methods of short-term driving behavior, using K-nearest neighbor (KNN), hierarchical clustering, and neural networks-self organizing maps as viable, partitioned clustering options for classifying behavioral features by the unsupervised machine learning method. Among these methods, several researchers used KNN to classify driving behavior [2], [28]. The efficiency of KNN in dealing with large datasets makes this method a perfect candidate for labeling unlabeled feature data. However, the output of KNN clustering failed to provide reasonable classification **[Figure 3(a)]**. The clusters that resulted from KNN were unable to represent explicit differences between two clusters. Increasing cluster size led to increased complexity in classification without proper explanation of individual cluster characteristics. Additionally, the clusters, specifically for highways and arterials, were incapable of addressing the impact of all three features in the classification process. Irrational division of traffic features resulting from KNN led us to examine the much simpler rule-based classification approach.

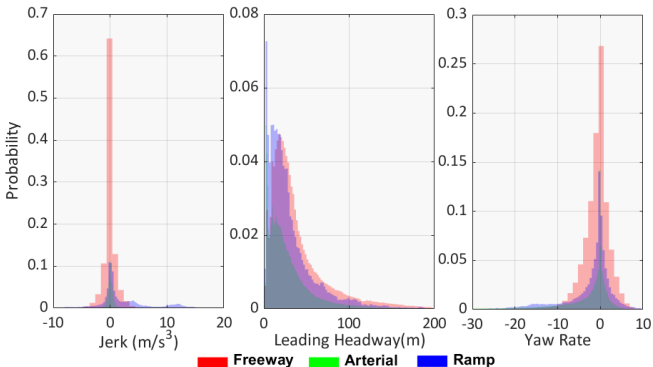

Fig. 2. Distribution of studied features on different road types.

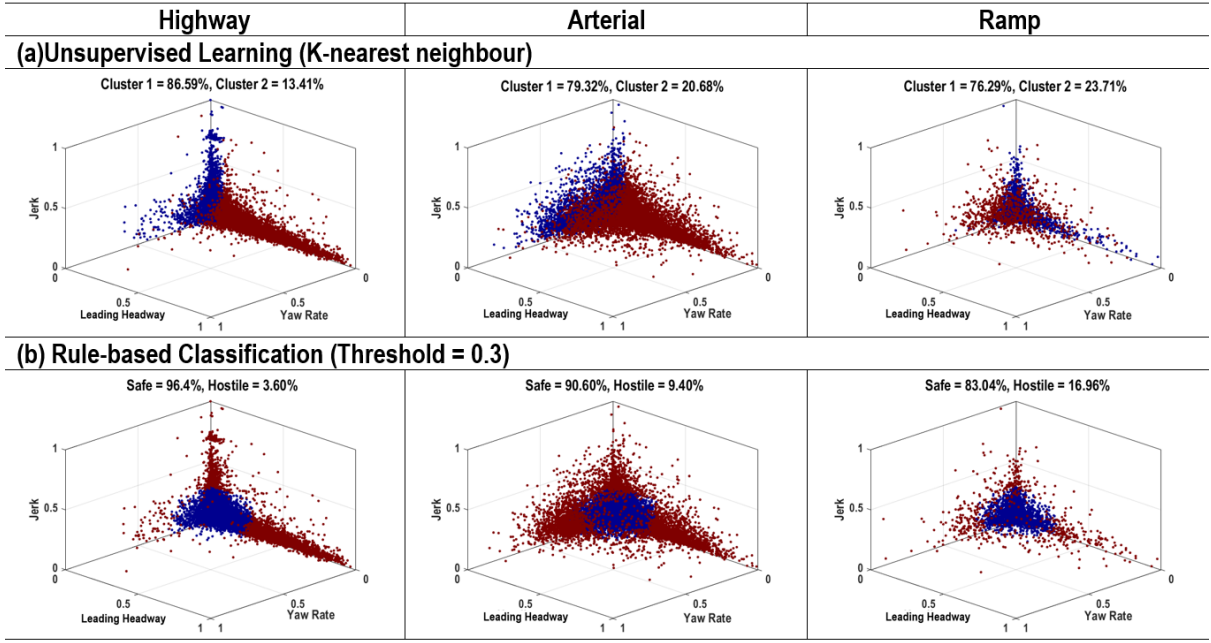

Fig. 3. Labelling scaled CoV values of studied features using (a) unsupervised learning and (b) rule-based classification methods.

Using the rule-based classification process, we chose a threshold value of scaled CoV to label driving decisions. If the scaled CoV value of any of our three features was higher than the threshold value, the driving behavior for that time window was labeled as 'Hostile Driving'. In the process of labelling traffic behavior, we explored different threshold values of CoV to identify the sensitivity of the threshold value. The results indicated that reducing threshold value of scaled CoV would lead to a higher share of 'Hostile Driving'.

Therefore, to remain on the conservative spectrum of behavior identification, we chose a small threshold value of

### TABLE I
### ALGORITHM FOR SHORT-TERM CLASSIFICATION LEARNER

1. Identified road type and CoV for corresponding road type within the time window $[(t - t_c),\ t]$
2. An unlabeled training set, $S_u^t = \left\{ (CoV'_{f=[jerk,ld.\ headw.,yaw\ rate]}(t) \right\}_{t=1}^T$,
   Here, T = number of training instances.
3. If $CoV'_{jerk}(t) \mid CoV'_{ld.\ headw.}(t) \mid CoV'_{yaw\ rate}(t) > threshold$ then Driving Behavior$_{short-term}[(t - t_c), t]$ = Hostile Driving

### TABLE II
### ALGORITHM FOR LONG-TERM CLASSIFICATION

1. Average road type specific hostility driving percentages and overall hostility driving percentages of studied driver
2. Average hostile driving shares from training dataset for each road type and overall trip
3. (a) Road-type Specific Classification
   If $\%_{Hostile\ Driving_R} < threshold_{lower} \times \%_{Hostile\ Driving_R}^{training}$
   then Driving Behavior$_{long-term}(R)$ = Calm Driver on road type 'R'
   Else if $threshold_{lower} \times \%_{Hostile\ Driving_R}^{training} \leq \%_{Hostile\ Driving_R} \leq threshold_{upper} \times \%_{Hostile\ Driving_R}^{training}$
   then Driving Behavior$_{long-term}(R)$ = Rational Driver on road type 'R'
   Else if $\%_{Hostile\ Driving_R}^{training} > threshold_{upper} \times \%_{Hostile\ Driving_R}^{training}$
   then Driving Behavior$_{long-term}(R)$ = Aggressive Driver on road type 'R'
   Here, R = road type (i.e. freeway, arterial, ramp).
3. (b) Overall Classification
   If $\sum_{R=\{freeway,arterial,ramp\}} \%_{Driving_R} \times \%_{Hostile\ Driving_R}) < threshold_{lower} \times \sum_{R=\{freeway,arterial,ramp\}} \%_{Driving_R}^{training} \times \%_{Hostile\ driving_R}^{training}$
   then Driving Behavior$_{long-term}$ = Calm Driver
   Else if $threshold_{lower} \times \sum_{R=\{freeway,arterial,ramp\}} \%_{Driving_R}^{training} \times \%_{Hostile\ Driving_R}^{training} \leq \sum_{R=\{freeway,arterial,ramp\}} \%_{Driving_R}^{training} \times \%_{Hostile\ Driving_R}) \leq threshold_{upper} \times \sum_{R=\{freeway,arterial,ramp\}} \%_{Driving_R}^{training} \times \%_{Hostile\ Driving_R}^{training}$
   then Driving Behavior$_{long-term}$ = Rational Driver
   Else if $\sum_{R=\{freeway,arterial,ramp\}} \%_{Driving_R} \times \%_{Hostile\ Driving_R}) > threshold_{upper} \times \sum_{R=\{freeway,arterial,ramp\}} \%_{Driving_R}^{training} \times \%_{Hostile\ Driving_R}^{training}$
   then Driving Behavior$_{long-term}$ = Aggressive Driver

scaled CoV (0.3) **[Figure 3(b)]**. Followed by the labeling process, we then executed several supervised classification learner methods (i.e. logistic regression, discriminant analysis, support vector machine, decision tree) over the labeled training data to identify the best classifying model. Among our explored models with 10-fold cross validation, the decision tree model provided the highest accuracy ($\sim$100%) in correctly classifying training data for all road types and, hence, was used as a short-term classifier. Table I summarizes the steps involved in labelling training datasets to enable subsequent classification.

While the selected threshold for classifying behavior was the same for all types of roads (i.e. freeway, arterial, ramp), the threshold value was applied on scaled CoV values of studied features, derived by balancing different ranges of feature values into a common unit. As illustrated earlier in Figure 2, the ranges of these features were significantly different with respect to different road classes. Hence, the same threshold value on scaled parameters resulted in different CoV values for different road classes. In the end, the behavior-classifying limit would remain the same for a specific feature on a specific road class and demonstrate a dynamic quality with changing road types as well as features.

We then used measured values of road type specific shares of 'Hostile Driving' on total driving instances to categorize long-term driving behavior. For instance, 9.40% of samples from total training data demonstrated 'Hostile Driving' behavior while driving through arterial roads. To recognize long-term driving behavior on arterials for a specific driver, we considered the accumulated classified (i.e. safe, hostile) driving history and compared the share of cumulative hostile driving decisions along arterial roads with training 'Hostile Driving' shares. For this analysis, we regarded 0.5 as the lower threshold and 1.0 as the upper threshold to classify long-term driving behavior into Calm, Rational, and Aggressive driving behavior. So, if the cumulative 'Hostile Driving' share along arterial of a driver was less than 4.7% (= 0.5 × 9.4%), then that driver was classified as a 'Calm Driver' on arterial roads. On the other hand, if the same share rose above 9.4%, then that driver was classified as an 'Aggressive Driver' on arterials. We followed a similar process to classify long-term behavior of drivers on other road types and total travel history. Table II describes the process of long-term behavior classification based on road types and overall driving history.

To provide further clarification of the long-term behavior classification process, a hypothetical scenario is presented here as illustration. Suppose a specific driver had made 30 trips, and the three feature values (i.e. jerk, yaw rate, leading headway) were collected, scaled, and stored according to the short-term behavior classification process. Then, the average hostile driving proportion of these 30 trips was measured for long-term behavior classification, using the three specified road types (i.e. freeway, arterial, ramp) as well as overall trips. By analyzing this road user's driving history of 30 trips, let us imagine that they showed average hostile driving behavior on freeway, arterial and ramps for 5.17%, 4.27% and 11.84% of the total driving time, respectively. We would find that the average hostile driving share for total trips to be 2.95% when the total number of trips were evaluated for driving behavior. Once these values were obtained from the driver's history, it would be compared with the stored road-specific and overall-average hostile driving shares of the training dataset. The average hostile driving shares of the training dataset would be 3.60%, 9.40%, 16.96%, and 5.79% for freeway, arterial, ramp, and total trip, respectively. Once calculated, these values would form the basis of road-type specific classification by comparing the driver's hostile share with the training datasets hostile share. In our example, this driver's hostile share on freeway (5.14%) was found to more than 1.0 ×hostile share of training data on freeway (3.60%), therefore, the driver's long-term behavior, based on their driving history of 30 trips, had classified them as an 'Aggressive Driver' on freeways. Similarly, road-type specific, long-term classification would label this driver's behavior on arterial, ramp, and total trips as a 'Calm driver' [ 4.27% < 0.5 × 9.40%], 'Rational driver'[0.5 × 16.96% < 11.84% <

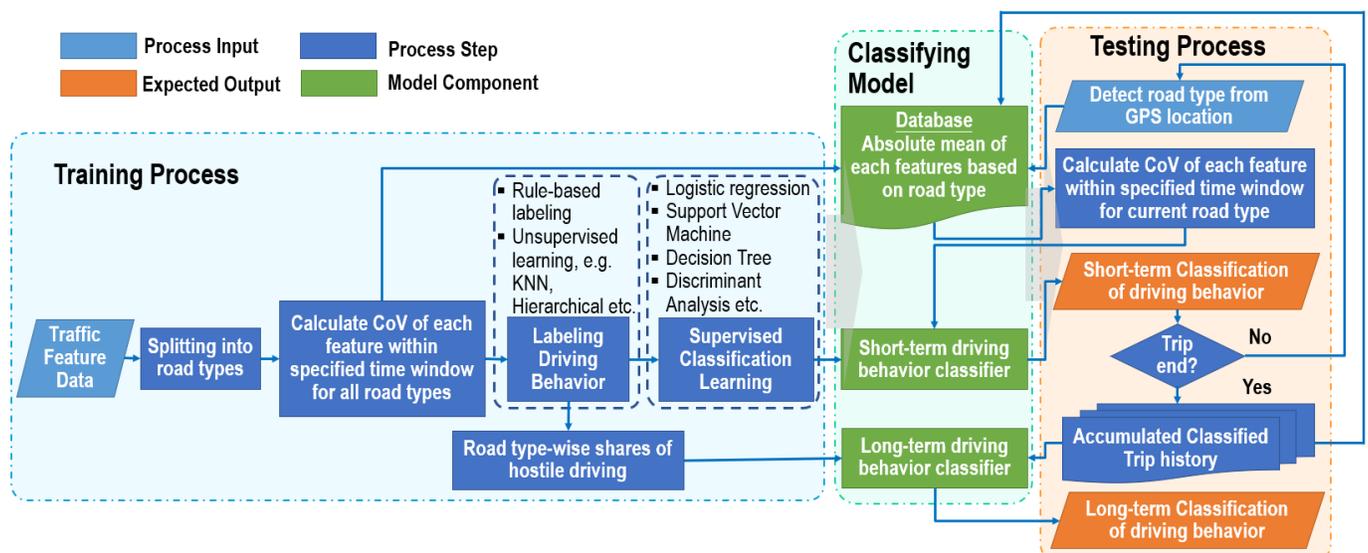

Fig. 4. Flow chart of the driving behavior classification algorithm.

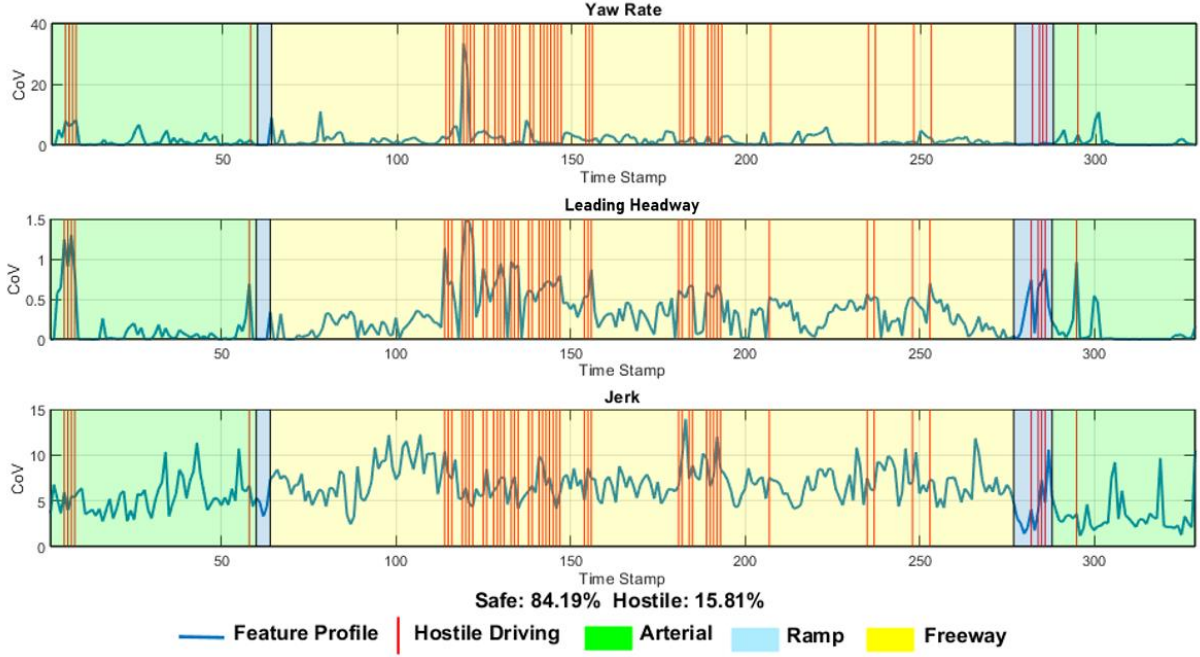

Fig. 5. Short-term driving behavior classification of a test trip.

$1.0 \times 16.96\%]$ and 'Rational driver' $[0.5 \times 5.79\% < 2.95\% < 1.0 \times 5.79\%]$, respectively. **Figure 4** presents the implemented classification algorithm in a flow chart in order to detail the progression of the behavior classification process.

## IV. PERFORMANCE EVALUATION

The generated classification models from the training data were executed on 'test trips' to classify driving behavior. To qualify as a 'test trip', we selected those with the highest number of datapoints (20% of training trips) among the remaining 110 trips on the database (except trips used for training purposes), which suggested that they were long and thus expected to contain the most diverse behavioral variations. We maintained the same time window of 5 sec (50 data points) to reshape classification features data. The proposed classifying model categorized the selected test trips for both short and long-term. The obtained hostility instances for the total trip of test trips varied between 1.45% to 18.53% with a mean of 5.67%. The short-term classification of total trips for a sample test trip is shown in **Figure 5**, which displays driving road types, the classification features' CoV profiles, and hostile driving instances during a 28min 23.7 sec long trip (341-time stamps). All 110 trips were categorized, with the short-term driving behavior classifier following the same process for specific road types and total trips.

Although the classifying model identified hostile driving behavior through longitudinal and lateral feature recognition, we had yet to test the precision of identified behavior. To do so, we took the velocity and acceleration profiles of each trip as explicit identifiers of hostile behavior. Then, mean velocities within a predetermined time window were measured and compared with the corresponding road type's speed limit. Subsequently, the time stamps with mean velocities higher than 10 miles above speed limits were labeled as 'Hostile Driving' instances. As a result, this classification method only used the speeding behavior of the driver. Several studies have selected speeding as the controlling feature of unusual driving instances identification [64]–[66]. A second process measured the acceleration range of each time window determined from the classification by acute acceleration change. Time stamps with an acceleration range higher than 2.5 m/s² were labeled as 'Hostile Driving' behavior. A similar approach to identifying unique driving events through acceleration variations had been used previously in numerous studies [25], [28], [65], [67]–[69]. Both explicit classification measures (i.e. classification by speeding, classification by acute acceleration change) were compared with the model classification output (i.e. short-term driving behavior classification) to evaluate the behavioral disparity identification capability of the proposed method. **Figure 6** presents a sample trip behavior classification using the aforementioned methods.

For the sample trip, a comparison of short-term behavior classification, from the generated classifying model with speeding-based classification, provided 87% accuracy. A similar comparison, with an acute acceleration-based classification, produced behavioral identification with 84% accuracy. Another analysis of speeding identification revealed that the proposed short-term classifying model accurately identified 19 out of 23 speeding instances as hostile driving behavior for the sample trip. Similarly, short-term classification identified 17 out 25 instances when compared to the acute acceleration change-based classification. The identification accuracy for all 110-test trips in comparison to the speeding-based classification was, on average, 86.31%, with a standard deviation of 9.84%. Likewise, the comparison with the acute acceleration change-based classification presented 87.92% average accuracy with 10.04% standard deviation.

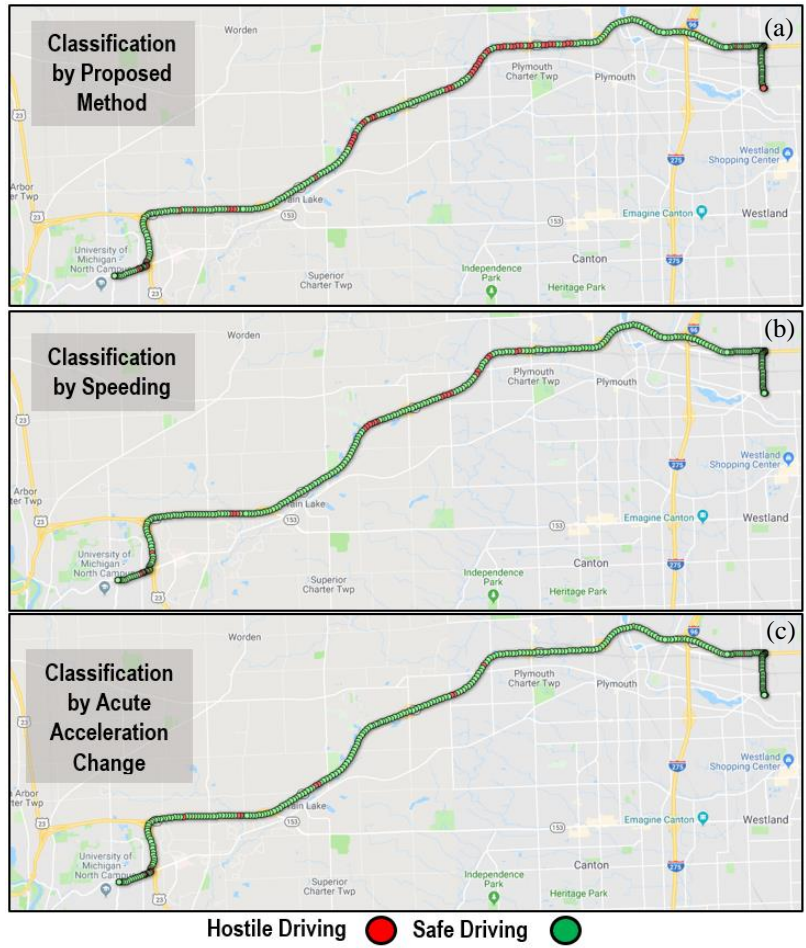

Fig. 6. Evaluation of (a) proposed classification method in comparison to (b) speeding-based classification, (c) acute acceleration change-based classification.

The short-term classification based on multiple driving features was further compared with the classification process proposed by Murphey et al. [1] to demonstrate the aptitude of the proposed methods in identifying behavioral extremity. Murphey et al. [1] proposed a single feature-based (i.e. jerk) classification of driving behavior into three groups (i.e. calm, normal, aggressive). The division of the groups were founded on threshold values of jerk profiles CoV (e.g. CoV of a time window < 0.5 then driving behavior = calm, 0.5< CoV of a time window< 1.0 then driving behavior = normal, 1.0 < CoV of a time window then driving behavior = aggressive). To measure the CoV, the average jerk value was measured on different road types and at different levels of service from 11 standard drive cycles. **Figure 7(a)** shows the classification of the sample trip by method in [1], and **Figure 7(b)** shows the classification of the same trip by method proposed in this paper. The average jerk for level of service C on a freeway, CD on arterial and ramps values were chosen to follow the jerk-based classification as these levels of services are usually expected in these road classes. Classification of the sample trip by the proposed method identified 13.09% of driving as hostile driving instances during the trips by analyzing three features, whereas classification by the method of Murphey et al. [1], identified 6.92% of driving as aggressive driving instances. Therefore, the additional features were capable of increasing the identification of hostile driving instances by just under 47%. Notably, the average jerk value used for calculating CoV was different for both methods, resulting in different jerk profile scales. Additionally, in contrast to the method in [1], the proposed method had a different threshold for different road types, generated by analyzing the training dataset.

To illustrate the long-term, behavior classification functionality of the proposed classifying process, the previously classified 110 test trips were presumed to be driven by the same driver at separate times. We found this assumption to be necessary since the demographic information on the dataset about the drivers making the trips was inaccessible. As such, it was impossible to link the dataset with a specific driver. Under this scenario, hostility shares on both specific road types and total trips were measured on the short-term classification. The hostility proportions of each trip were also compared with the training data's hostility proportions and classified into Calm, Rational, and Aggressive driving behavior by scaling training hostility shares with the lower threshold (0.5) and upper threshold (1.0). Long-term categorization was performed by measuring the moving averages of hostility shares (including all previous trips) and by matching that measurement with the hostility limits (<0.5: Calm, 0.5- 1.0: Rational, >1.0: Aggressive) of three groups (i.e. calm, rational, aggressive). **Figure 8** illustrates both types of test trip classification for specific road types as well as for the total trip. Each blue dot on the plots of **Figure 8** represent the hostility proportion of each trip that could be utilized to perform short-term classification. The red curve on the plots portrays the progression of driver behavior by taking all previous trips into account (moving averages of blue dots). Different color patches (i.e. yellow, green, red) on the plots illustrate the boundary regions of specific behavioral classes (i.e. calm driver, rational driver, aggressive driver). While individual trip hostility fluctuated frequently, the behavioral progression on long-term was relatively stable and changed gradually over time.

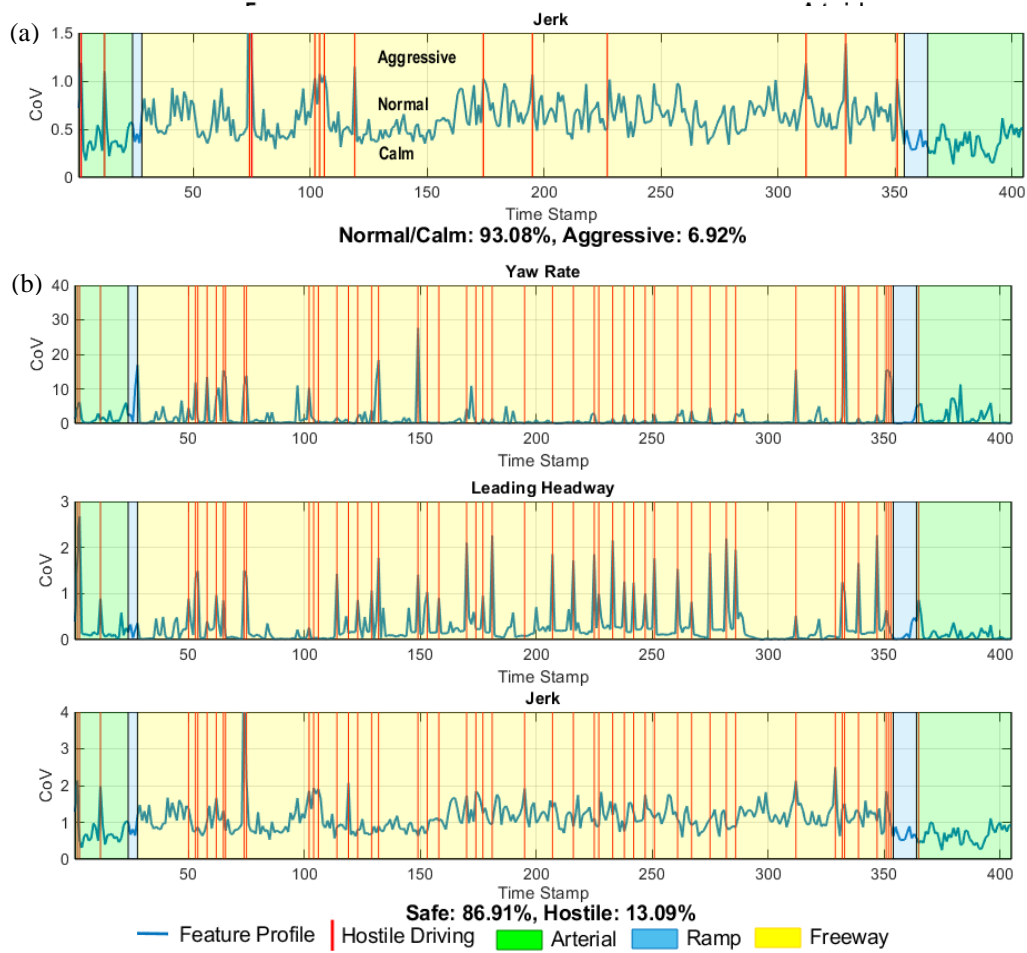

Fig. 7. Behavioral classification of a sample trip by (a) analyzing jerk feature and (b) analyzing multiple driving features.

The proposed method of long-term classification was capable of identifying the changing patterns of driving habits for the number of total trips and road type specific habits. In **Figure 8**, the total trip hostilities of the accumulated trips were highly weighted towards to freeway hostility, which suggests that the largest portion were traversed though freeways. Moreover, the comparison between freeway and arterial hostility share demonstrated higher long-term behavioral variability on arterial roads (standard deviation = 1.48%) than on freeways (standard deviation = 0.84%). The paired sample t-test on long-term freeway and arterial hostility showed significantly lower hostility on freeways at 95% confidence interval (t-score = 29.557, p-value < 0.001). The obtained comparison result did not necessarily mean that the driver was more aggressive on arterials than freeways, because the classifying threshold for freeways was different. As a result, the long-term behavior on arterials graduated from 'aggressive' to 'rational', even with higher hostility than freeways. Since ramp road-types had a relatively low share (2.5% in average) on total test trips, the influence of long-term ramp hostility on total trip hostility was discarded for comparison.

We further analyzed the identified hostility of 110 test trips to reveal short-term behavioral distribution on different road types. As shown in **Figure 9**, the hostility behavior was different from one road type to another. For instance, freeway hostility was skewed towards the origin, with the highest proportion lying between 2.5-5.0%. This skewness towards lower hostility could be explained by the fact that drivers, in general, tend to operate with less variations in control while driving on freeways (Figure 1(c)). Whereas, the probability distribution of arterial hostility was relatively balanced over a larger range of hostility (0-27.5%). The drivers had to experience more frequent disruptions, due to geometry, traffic control measures etc., while driving through arterials that could result in such diverse hostility patterns on arterials. Similarly, ramp hostility showed a central tendency towards the median. Since ramps perform as connecting links between freeways and arterials, we expected the hostility pattern in this transitional phase to be influenced by both road types' distribution. We performed a paired, two-sample t-test between the measured hostility ranges to identify significant dissimilarity in behavior on distinct road types. The results of the t-test showed that the hostility behavior on a specific road type was significantly different from other road-types with a 99% confidence level.

## V. FUTURE EXTENSIONS

Since the major motivation of behavior classification lies in persuading drivers to maintain safe driving patterns, providing real-time feedback on driving style is imperative in order to harness its benefits. With the assistance of CV technology and smartphones, driving behavior information can be conveyed to drivers through a user-friendly ADAS interface, designed to easily communicate both short-term and long-term behavior classification information **(Figure 10)**. Verbal and visual warnings on the ADAS interface can announce detected hostile behavior through short-term classification. **Figure 10(a)** provides an interface design for this purpose. The yellow circle in the middle starts to blink once hostile driving behavior is detected, thus presenting the driver with the visual warning. The system can deliver an auditory warning (indicated by the alarm sign on the picture) in tandem. In addition, the same interface can provide other information as part of the system design. This real-time warning system is assumed to induce cautiousness in drivers and, hence, promote safe driving behavior. At the same time, options to personalize the hostility threshold for different road-types and overall trip can be provided on the developed application, giving freedom to users to define their own hostility perception.

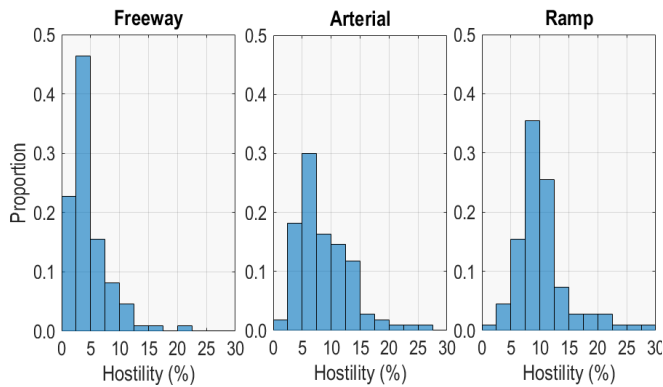

Fig. 9. Hostility distribution for test trips on different road types

In addition to real-time response, driving habits of individual drivers can be tracked through long-term classification, enabled by storing classified trip characteristics in a database. Previous classified trip history could be analyzed through the long-term classifier and displayed on a convenient interface to identify both road-type specific and overall driving habits **(Figure 10(b))**. The left most dial on **Figure 10(b)**, shows the overall long-term behavior classification from the trips within the time range, where the yellow region indicates 'Calm', green for 'Rational', and red for 'Aggressive'. The indicator arm of the dial gauge lies within the yellow and green region, which suggests the driver behavior falls within Calm and Rational driving behavior. The other three gauges in **Figure 10(b)** show road type specific long-term driving behavior and trip shares on each road type (value at the bottom right corner of each gauge).

Detected long-term driving behavior can assist road traffic operation and safety authorities, insurance companies, and other associated organizations to offer incentives for 'Rational Drivers' as well as to impose penalties on 'Aggressive Drivers' as a means to promote safe driving on roadways. As part of this continued research, we plan to develop a smartphone application to detect and communicate driving behavior information to drivers in real time. Furthermore, the application will store both short-term and long-term driving history as well as analyze the effects of ADAS on drivers' behavior and habits. The goal of the analysis will be to determine the capability of ADAS in bringing paradigmatic shifts in driving behavior.

## VI. CONCLUDING REMARKS

This paper presents a simple, efficient, and adaptable driving behavior classification technique developed by analyzing both longitudinal and lateral driving features collected though CV technology from real-world trips. The thresholds of the proposed classification method can be modified to accommodate transportation, motoring, and roadways authorities' purposes and requirements. By considering bidirectional features of driving, the proposed method has greater aptitude in sensing unsafe driving behavior compared to singular feature-based classification methods. This paper has taken a unique approach by distinguishing between driving behavior and driving habit as well as classifying drivers' behavior from both behavioral and habitual contexts. We worked with the concept of instantaneous behavior classification and used that information to categorize drivers' driving habits. Authorities considering the uses of behavior classification are not only interested in current responses but also in driving style, with the aim of recognizing safety hazards caused by those drivers and the extent of safety risk taken by allowing them to drive. Our study, given its scope, would help facilitate their decision-making concerning rewards and penalties for driving behavior.

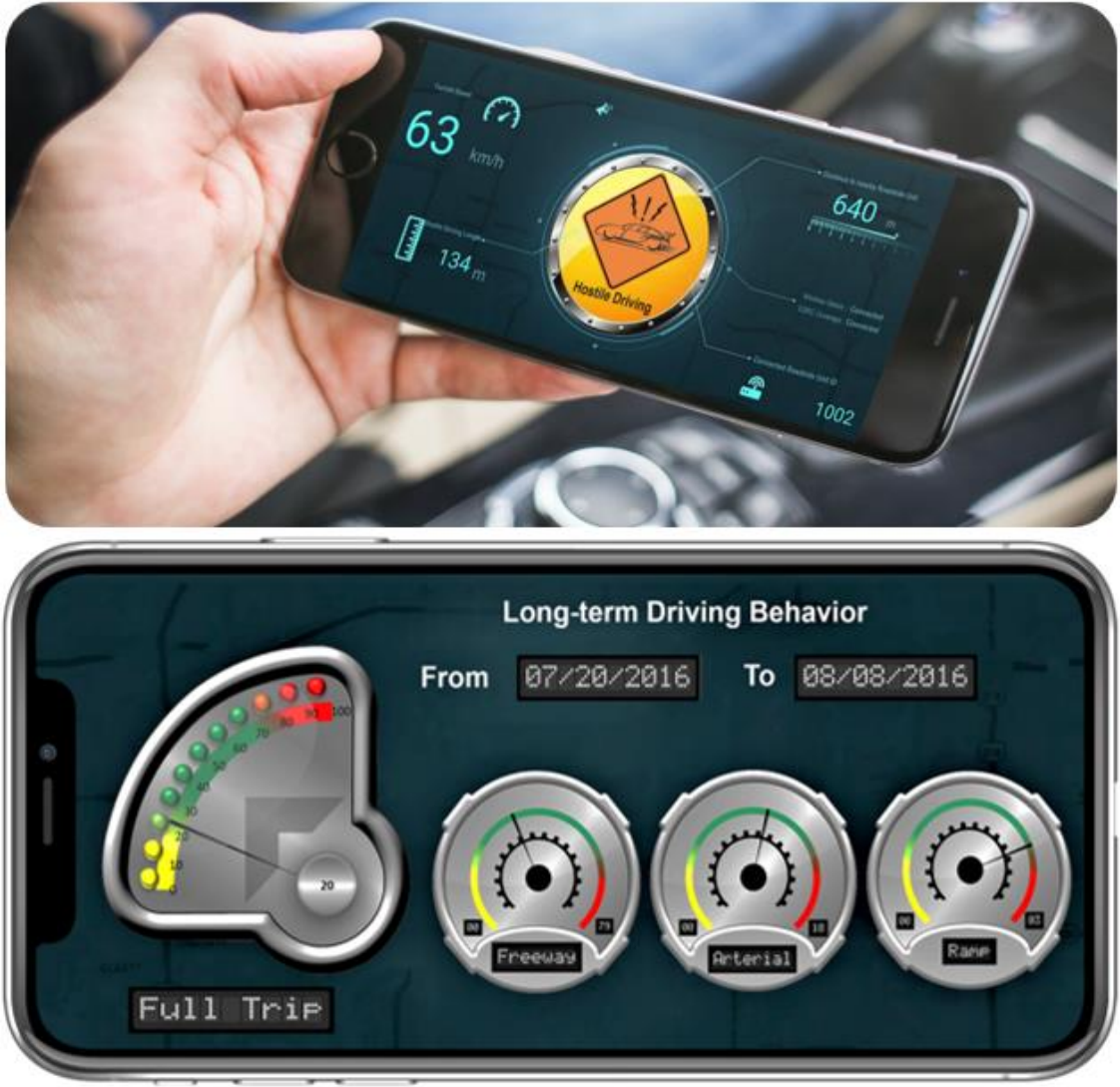

Fig. 10. Abstract ADAS interface for communicating (a) real-time warning, (b) long-term behavioral information to drivers

While we have limited our research to three distinct features in the form of continuous variables, in order to illustrate longitudinal and lateral decisions, other features can also distinguish characteristic identifiers. Furthermore, we analyzed partial datasets of a larger SPMD database to demonstrate the classification technique. Since the primary aim of the study is to propose and present a simplified classification technique, we have set aside the potential bias of the analyzed datasets. In brief, this study is an attempt to gain insight into driving behavior and habits though a simple categorization process that considers bidirectional control decisions. Furthermore, our study offers the possibility for extension through the development of ADAS and through the identification of its impact on modifying driving behavior and habits.

## ACKNOWLEDGMENT

This research work was supported by the Natural Sciences and Engineering Research Council (NSERC) of Canada, City of Edmonton, and Transport Canada. The contents of this paper reflect the views of the authors who are responsible for the facts and the accuracy of the data presented herein. The data that supports the findings of this study is available from the corresponding author, upon reasonable request. The contents do not necessarily reflect the official views or policies of the City of Edmonton or Transport Canada. This paper does not constitute a standard, specification, or regulation. The authors declare that there is no conflict of interest regarding the publication of this paper.